\documentclass[twocolumn,trackchanges]{aastex7}
\usepackage{floatrow}
\usepackage{float} 
\begin{document}

\title{Evaluating the Limits of Rotation Period Recovery through Gyrochronology Criteria}

\author[0000-0002-4697-6565]{Mariel Lares-Martiz}
\affiliation{Embry-Riddle Aeronautical University \\
1 Aerospace Blvd., Daytona Beach, FL 32114, USA}
\email[show]{laresm@erau.edu} 

\author[0000-0002-8541-9921]{Terry D. Oswalt}
\affiliation{Embry-Riddle Aeronautical University \\
1 Aerospace Blvd., Daytona Beach, FL 32114, USA}
\email[]{oswaltt1@erau.edu} 

\author[0000-0002-1988-143X]{Derek L. Buzasi}
\affiliation{University of Chicago \\
5801 S. Ellis Ave., Chicago, IL 60637, USA } 
\email[]{dbuzasi@uchicago.edu}

\author[]{Kylie R. Boyer}
\affiliation{Embry-Riddle Aeronautical University \\
1 Aerospace Blvd., Daytona Beach, FL 32114, USA}
\email[]{BOYERK@my.erau.edu} 

\author[]{Luca Guida}
\affiliation{Embry-Riddle Aeronautical University \\
1 Aerospace Blvd., Daytona Beach, FL 32114, USA}
\email[]{GUIDAL@my.erau.edu} 

\author[]{Ryan J. Reynolds}
\affiliation{Embry-Riddle Aeronautical University \\
1 Aerospace Blvd., Daytona Beach, FL 32114, USA}
\email[]{REYNOR20@my.erau.edu}

\begin{abstract}

Contamination from nearby sources often compromises stellar rotation periods derived from photometric light curves, particularly in data with large pixel scales such as TESS. This problem is compounded when both the target and contaminant are intrinsically variable, a scenario that challenges deblending algorithms, which often assume constant contaminants. We assess the reliability of rotation period detections using wide binary systems, whose components share a common age and rotational history. By applying gyrochronology constraints, we identify period combinations that yield consistent ages between components, helping to isolate true rotation signals. Simulating blends with degraded Kepler data, our method recovers correct rotation periods with an 88\% success rate for periods $<12$ days, where TESS detections are most reliable. Applying this framework to nearly 300 wide binaries observed by TESS, we find that despite significant contamination, a subset of pairs shows consistent gyrochronological ages. We establish a practical detection threshold for TESS blended observations, finding that periods shorter than $\sim8$ days are reliably recovered, while those longer than $\sim10$ days become significantly more challenging and often remain unresolved. As expected, rotation periods are more often recovered when the highest-amplitude periodogram peak is linked to the brighter star and the second to the dimmer star, although many cases deviate from this pattern, indicating it cannot always be assumed. Our results highlight the limitations of standard deblending methods and demonstrate that astrophysical constraints, such as gyrochronology, provide a valuable tool for extracting reliable rotation periods from complex photometric blends.

\end{abstract}



\section{Introduction}\label{sec: Intro}

Gyrochronology is a tool for determining stellar ages that has been available for over five decades since \cite{Skumanich1972} showed the rotation period-age relationship. Subsequent studies of open clusters have shown that the rotation–age relation depends strongly on stellar mass, reflecting differences in the rates of angular momentum loss among stars of different masses \citep{Rebull2016, Douglas2017, Douglas2019, Curtis2019, Curtis2020, Gruner2023M67}.  Using several open clusters and a large sample of wide binaries (WBs) in the Kepler and K2 fields, \cite{Gruner2023} demonstrated that WBs, implying field stars, followed the gyrochronology paradigm as well as clusters.  Although no physical models have been devised to date that match these observations, and gyrochronology remains a purely empirical method, it can still be used to estimate ages for a vast range of stars. 

Implementing a gyrochronology tool to date stars is relatively easy; all that is needed is a color index (or any other mass proxy) and a reliable stellar rotation period. The photometric data in Gaia EDR3 provides exceptional color precision, with systematic errors kept below the $1\%$ level across the entire magnitude and color range \citep{Brown2021}. However, the situation for rotation periods is considerably more challenging. 

Several factors significantly hinder the reliability of rotation periods. For example, misidentification of signals that mimic rotational variability (pulsations, eclipses, or harmonics) \citep{santos2017,Santos2019,Santos2021} and limitations of photometric data (TESS’s large pixel size, limited duration of each observing sector, instrumental systematics) introduce artifacts or obscure true rotational signals \citep{Claytor2022}. Even though some recent studies assert that it is possible to detect long rotation periods with TESS data \citep{Hattori2025}, other studies demonstrate that rotation periods recovered from single-sector TESS observations become increasingly unreliable beyond $\sim10$ days, reflecting both insufficient cycle coverage and rising noise levels \citep{Boyle2025}. More advanced analytical techniques and algorithms are required to distinguish genuine rotational signatures reliably.

Currently, with more than seven cycles of TESS observations available, several tools have been developed that address these obstacles \citep{Hedges2020, Hedges2021, Han2023, Binks2024} and the limits of rotation period determinations \citep{Lu2020, Boyle2025}. Especially noteworthy is TESS\_localize \citep{Higgins2023}, a sophisticated algorithm designed to identify the origin of the signal within the Target Pixel File (TPF) data. This tool leverages the high-frequency sensitivity of TESS data to mitigate its poor spatial resolution. However, most of these deblending methods require or assume that contaminating sources are \textit{non‐variable}, which limits their effectiveness in systems where both stars exhibit intrinsic variability. 

\begin{figure}[h]
    \centering
    \includegraphics[width=\linewidth]{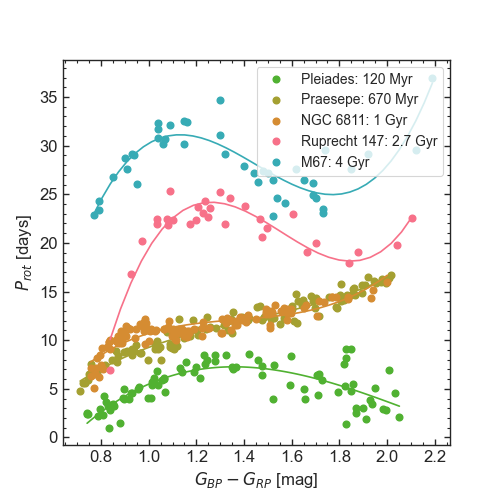}
    \caption{CPD for several open cluster's most recent data.}
    \label{fig: openclusters}
\end{figure}

Here, we present work using the well-established empirical gyrochronology paradigm, i.e., for stars in the 3800-6200 K effective temperature range \citep{Bouma2023} to determine the reliability of rotation periods. Rather than devising a new method, we aim to explore the boundaries of what can be achieved when applying gyrochronology criteria to potentially contaminated light curves. We take advantage of the properties of WBs, especially the strong constraint that components of each pair are coeval. We analyze a sample of WBs observed by TESS. Theoretically, both components of a wide binary should follow the same gyrochronology relation, meaning they should lie along a line of constant age (``gyrochrone") in the Color–Period Diagram (CPD; see Figure~\ref{fig: openclusters}). 

Since the dominant source of scatter in a CPD arises from uncertainties in the rotation period, we hypothesize that if one component deviates significantly from the expected gyrochrone, its apparent rotation period is likely to be inaccurate. In this work, the gyrochrones are derived from empirical fits to open cluster data, which naturally include observed phenomena such as the spin-down region associated with weakened magnetic braking. Therefore, our method takes these features into account.

This paper is organized as follows. In Section~\ref{sec: method}, we introduce the coevality criterion (CC) approach for WBs, outlining how gyrochronology constraints can be used to identify the most likely period combinations in blended WB light curves. Section~\ref{sec: degraded-test} presents tests using simulated blends of Kepler light curves to evaluate the method’s performance under controlled conditions, where the true stellar rotation periods are known. In Section~\ref{sec: results}, we apply the approach to a sample of WBs observed by TESS, highlighting the challenges of contamination and the conditions under which accurate period recovery is still achievable. Finally, Section~\ref{sec: discussion} summarizes our main conclusions and discusses the broader implications of these results for stellar rotation studies, emphasizing the value of gyrochronology as a tool for validating rotation periods in the TESS era.

\section{The Rotation Period Agreement Approach for Wide Binaries} \label{sec: method}

To evaluate the reliability of stellar rotation periods detected in TESS light curves, we compare them to the established empirical relationship between stellar rotation period and age, known as gyrochronology. Instead of using theoretical models, we utilize real observational data from open clusters, which currently provide the most accurate representation of the stellar age-rotation period relationship. We chose five well-studied open clusters: Pleiades \citep{Rebull2016}, Praesepe \citep{Douglas2017,Douglas2019}, NGC 6811 \citep{Curtis2019}, Ruprecht 147 \citep{Curtis2020}, and M67 \citep{Gruner2023M67} , each with a well-established age and rotation period distribution.

\begin{deluxetable}{lc}
\tablecaption{Two-sigma ($2\sigma$) deviations around the gyrochrones for each open cluster used as reference when applying the CC. These thresholds define the allowed deviation in rotation period from the gyrochrones for a pair to be considered consistent with a common gyro age. \label{tab:cluster_sigmas}}
\tablehead{
\colhead{Open Cluster} & \colhead{2$\sigma$ Deviation [days]}
}
\startdata
Pleiades     & 2.73 \\
Praesepe     & 1.32 \\
NGC 6811     & 1.64 \\
Ruprecht 147 & 2.78 \\
M67          & 3.56 \\
\enddata
\end{deluxetable}

The first step in this comparison involves performing a third-degree polynomial fit to the observed data for each cluster (see Figure~\ref{fig: openclusters}). These fits trace a continuous relationship between age and rotation period, namely a \textit{gyrochrone}, based on stellar observations. The polynomial fit from these clusters provides a robust reference for the expected rotation periods for stars of several known ages. The standard deviations listed in Table~\ref{tab:cluster_sigmas} represent the $2\sigma$ thresholds around each cluster's gyrochrone and define the allowable deviation in rotation period for a pair to be considered consistent with a common gyro age. The choice of a third-degree polynomial was made because higher-order fits did not significantly improve the correlation coefficient, allowing us to keep the model simpler. The $2\sigma$ threshold was chosen to reflect the typical scatter of data points in the open clusters. Later, in Section \ref{sec: degraded-test}, we confirm that the combination of a third-degree polynomial and a $2\sigma$ threshold achieves an optimal balance of model simplicity, reliability, and completeness.

\subsection{Coevality Criterion}\label{subsec:proximity}

We hypothesize that stars in WBs, which are expected to share a common age and metallicity, should exhibit rotation periods that align with the gyrochronology paradigm. To test this hypothesis, we apply a Coevality Criterion (CC): stars whose rotation periods fall within two sigma ($2\sigma$) around the polynomial fit for the relevant open clusters are considered to have reliable period determinations. This criterion assumes that the stars in WBs are sufficiently similar in age, so their rotation periods should be consistent with the cluster's gyrochronology curve. This is a reasonable assumption supported by \cite{Gruner2023} 's findings, which show that rotation-based age estimates usually are consistent for both components in WBs. While proximity to a cluster’s gyrochrone does \textit{not} imply cluster membership, it provides a useful reference for identifying coeval systems.

However, in some cases, discrepancies arise from physical scenarios such as unresolved hierarchical systems, orbital motion, tidal interactions, or mass exchange dominating the apparent rotation period.  They may also reflect systematic instrumental biases or contamination of the observational data. Since it is not possible to determine \textit{a priori} which of these scenarios is responsible for the disagreement, we treat all such cases uniformly as ``disagreeing pairs", regardless of the underlying cause, and investigate whether any distinguishing patterns emerge within this group. As an exception, we exclude from our analysis pairs in which both components exhibit rotation periods shorter than one day because, in such cases, rotation is unlikely to be the cause of variability. 

We validated our CC approach by showing that young ``agreeing WBs"—those whose rotation periods are consistent with the gyrochrones of open clusters—also have independently estimated young ages based on their galactic kinematics (see Section \ref{subsec:kinematics}). This consistency indicates that agreement with a cluster gyrochrone is a reliable indicator of the correct period identification, even when the data may be affected by contamination or uncertainties in period determination.

\begin{figure}[!ht]
    \centering
    \includegraphics[width=1.0\linewidth]{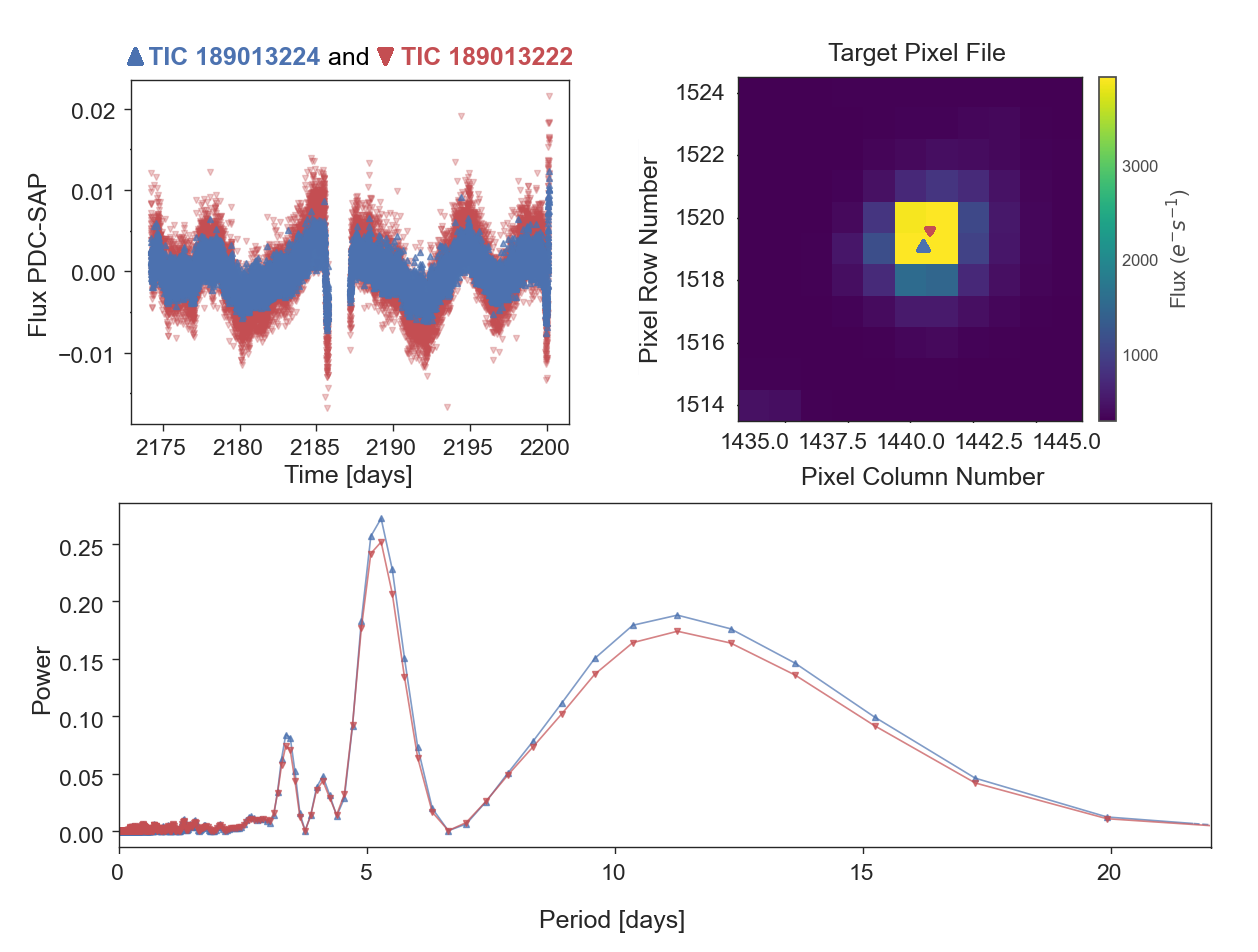}
    \caption{Example of blending in the pair TIC 189013224 and TIC 189013222. Top left: PDC–SAP light curves of TIC 189013224 (blue) and TIC 189013222 (red), showing their blended variability over the $\sim25~day$ interval. Top right: Corresponding TESS target-pixel file (TPF) for the blend, with the approximate centroids of TIC 189013224 marked by a blue triangle and TIC 189013222 by a red inverted triangle. Bottom: nearly identical Lomb–Scargle periodograms computed for each component’s extracted flux.}
    \label{fig:fig:ex2_tpf+lc+ls blend}
\end{figure}

\subsubsection{Blends} \label{subsubsec:blends}

In the CPD, blended WB sources often manifest as nearly horizontal alignments, where the two stars appear to share the same rotation period despite having significantly different colors (or spectral types). This pattern often arises due to flux contamination, where the light curves of both stars contribute nearly equally to the detected variability, leading to the spurious identification of a single dominant period (See Figure \ref{fig:fig:ex2_tpf+lc+ls blend}). While in some cases, this alignment coincides with the expected rotation period distribution of open clusters—particularly in younger clusters like the Pleiades, where the gyrochronology relation follows a relatively flat trend—such occurrences are improbable among a broad range of ages and stellar types. In general, as can be seen in Figure \ref{fig: openclusters}, it is unlikely for two stars of substantially different spectral types to exhibit nearly identical periods unless the detection is affected by blending.

For this reason, we do not consider such cases to be correct period determinations because they violate the CC. For both blended sources and systems where the identified periods deviate from expected gyrochrones, we implemented an additional criterion:  recovered periods must be physically consistent with stellar evolution models as well as gyrochronology expectations.  This helps filter out cases where contamination or systematic effects have led to incorrect period identifications.

\begin{figure*}[!ht] 
    \centering
    \includegraphics[width=\textwidth,height=9cm]{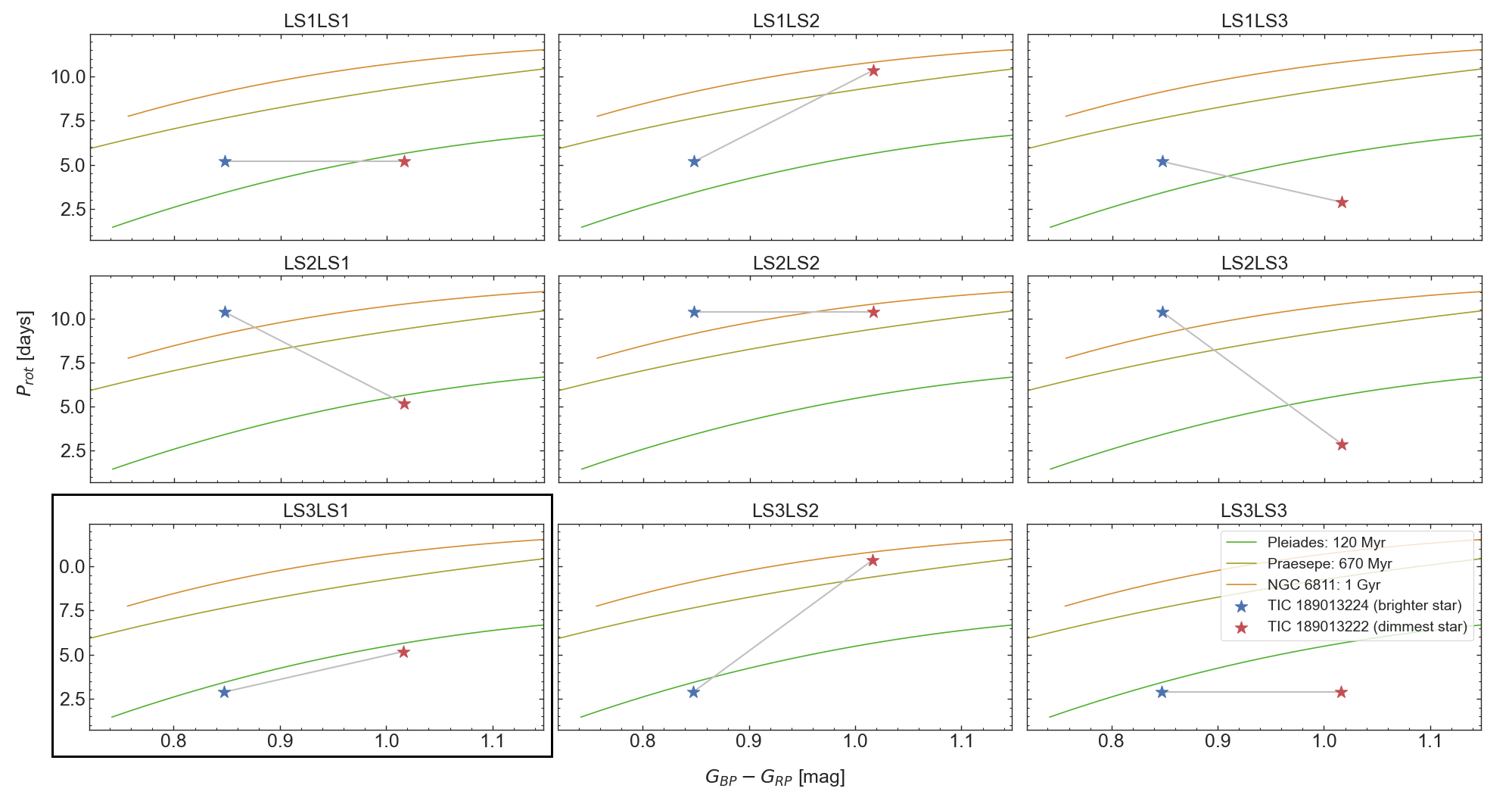}
    \caption{CPD showing every possible pairing of the three dominant Lomb–Scargle (LS) peaks from the blended light curves of TIC 189013224 and TIC 189013222. Each panel is labeled “LSaSb,” where “a” denotes the rank of the LS periodogram peak assigned to Component 1 and “b” denotes the rank assigned to Component 2 (for example, “LS1LS2” means the first‐ranked peak for the primary star and the second‐ranked peak for the secondary). The panel outlined in black highlights the specific combination of peaks that best satisfies the CC.}
    \label{fig:fig:ex1_allcombis}
\end{figure*}

\subsection{Reconciliation step}\label{subsec:proximity}

For each WB whose initial rotation period determinations did not satisfy the CC we selected the three most significant peaks from each WB component's periodogram.  We then evaluated all possible combinations of peaks between the two components, searching for the two periods that placed both components within $2\sigma$ of the gyrochrone of a given open cluster data. We adopted this threshold during parameter calibration because it falls within the open clusters data’s natural dispersion yet remains permissive enough to provide a substantial sample. If a combination of two peaks satisfied the $2\sigma$ proximity criterion, the corresponding periods were considered consistent with a common gyrochronological age (see Figure~ \ref{fig:fig:ex1_allcombis}).

For ambiguous cases where multiple peak combinations satisfied the CC, we ordered the combinations by their combined LS power and then select the top-ranked pairing (i.e., the first entry in the list). This rule yields a variety of combinations (e.g., LS1LS2, LS1LS3, LS2LS3) and is empirically justified and discussed in Section \ref{subsec:simul-results}.

\section{Testing Period Recovery with Simulated Blends } \label{sec: degraded-test}

The $\sim 4$" pixel size of Kepler data is much less subject to stellar blends than TESS $\sim 21$" pixels. By artificially combining the fluxes of real Kepler light curves with known rotation periods, we investigated how blending affects period detection. This analysis provides insight into the biases introduced by contamination and serves as a benchmark for interpreting rotation periods in blended TESS observations.

For our analysis, we used the sample of WBs identified by \cite{Gruner2023}, which includes systems with well-determined rotation periods from Kepler and K2 light curves. \cite{Gruner2023} applied rigorous selection criteria to ensure the reliability of the detected periods, using multiple independent diagnostics to confirm periodicity and to mitigate the effects of spot evolution, differential rotation, and instrumental systematics. As a result, rotation periods for all binary components in their sample are highly robust, making this dataset ideal for evaluating the CC we devised for dealing with TESS data.

From Gruner’s sample of 304 wide binary pairs, we applied a series of selection criteria to ensure a clean comparison baseline. First, we excluded any pairs lacking color information or with Gaia color index outside $ 0.75 < (G_{BP}-G_{RP}) < 2.0$, which corresponds to the gyrochronology regime. We also removed systems in which either component was flagged as problematic or annotated with cautionary notes in Gruner's catalog. Finally, we retained only those systems that Gruner classified as “S”, where both stars fall on the slow-rotator sequence \citep{Barnes2003} at a consistent gyrochronology age, ensuring the selection of physically meaningful, coeval systems. This final set of 89 pairs constitutes our “ground truth” sample for evaluating rotation period recovery.

\subsection{Simulated Blended Light Curves and Periodogram Analysis} \label{subsec:simul-lc}

To replicate the effects of blending, we selected each pair in the \cite{Gruner2023} WB sample and combined their component stars' fluxes, mimicking the frequent TESS scenario where two unresolved sources contribute to the observed variability. This was achieved by summing the normalized fluxes of each pair, scaled by a weighting factor that considers the percentage of total flux contributed by each star within a given pixel. Specifically, we computed the fraction of light contributed by the secondary star assuming no prior knowledge of the blending fraction. This was estimated directly from the magnitude difference between components as:
\begin{equation}
   f=\frac{F_{c2}}{F_{c1}+F_{c2}}=\frac{10^{-0.4\Delta{Gmag}}}{1+10^{-0.4\Delta{Gmag}}},
\end{equation}
where c1 and c2 corresponds to components 1 and 2, $\Delta{G{mag}} = G{mag_{c2}}-G{mag_{c1}}$, where magnitudes are extracted from the Gaia catalog. We used the magnitudes to compute the flux ratio and recover the true photometric contribution of each component, since the targets were observed in different quarters and had to be stitched together, resulting in a normalized light curve that does not preserve the absolute flux scale. This approach allows us to simulate realistic cases where one star dominates the signal while the other introduces secondary variability, as observed in blended sources from TESS (see Figure \ref{fig:fig:ex2_tpf+lc+ls blend}).

To further approximate TESS observational limitations, we degraded the Kepler light curves by introducing additional noise consistent with the lower photometric precision of TESS full-frame images. We added Gaussian white noise with a standard deviation of 0.5\%, representative of the photometric precision typical for stars with TESS magnitudes around 14.5–15 \citep{Sullivan2015}. This level was chosen to approximate the point-to-point scatter observed in real FFI light curves of moderately faint targets \citep{Vanderspek2018}. While actual TESS light curves contain red (or pink) noise components due to instrumental systematics and scattered light, we chose to use white noise at this level to ensure our simulations remain conservative. Most of our targets are brighter than T = 15, thus expected to exhibit lower intrinsic noise, so this approach provides a reasonable upper bound. Finally, we applied Gaussian smoothing to mimic additional instrumental trends.

We prewhitened the first 10 dominant frequencies from the simulated blended light curves and then ensured that the chosen frequencies were separated from one another by at least $0.003~d^{-1}$. This threshold, more stringent than the frequency resolution of the periodogram, is used to avoid selecting peaks that might arise due to differential rotation or a distribution of spots, thus ensuring that the identified signals correspond to distinct physical periods.

By applying our CC to these simulated TESS-like light curves, we evaluated the conditions under which the correct rotation periods can be retrieved, whether spurious signals exist, and how contamination affects the dominant peaks in the periodogram. This analysis provides a controlled framework to test the potential for distinguishing true stellar rotation periods from blending-induced artifacts.

\subsection{Results from the Blending Simulation: Trends in Recovered Periods} \label{subsec:simul-results}

One of the first things we noticed is that pairs with exactly one combination passing the gyrochronology constraint are strongly skewed toward longer (slower) rotation periods—presumably because gyrochrones are generally more widely separated at greater periods. To mitigate this and to improve completeness, we added a tie-breaker selection rule in cases with multiple valid pairings: we ranked all coevality-passing combinations by their combined LS power and adopted the top-ranked pairing (i.e., the first list entry). This tie-breaker is neither \textit{ad hoc} nor circular: among unambiguous pairs, the highest-power solution already dominates, and although it introduces a mild bias toward dominant peaks from the brighter component, it substantially increases our usable sample. We applied this rule uniformly across all 89 simulated blended pairs.

\begin{figure}
    \centering
    \includegraphics[width=1.0\linewidth]{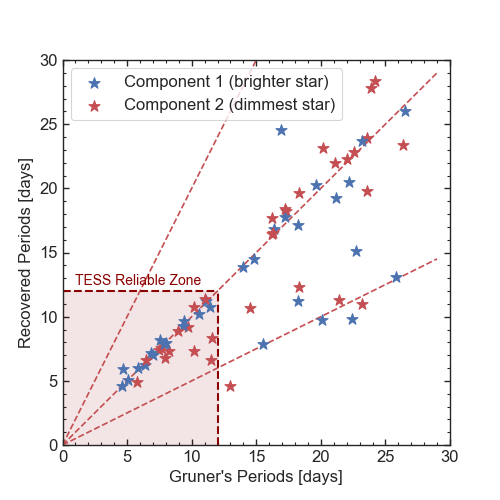}
    \caption{ Comparison between the recovered rotation periods from simulated blends and the original periods reported by \cite{Gruner2023} for both components in WBs. Red dashed lines mark the 2:1, 1:1, and 0.5:1 period ratios. The shaded region, bounded by dashed lines at 12 days, highlights the TESS Reliable Zone. In this period range, TESS is most sensitive and reliable for detecting rotation signals due to its 27-day observing window per sector. Within this region, our method correctly recovers 88\% of the original periods, demonstrating strong performance where TESS is most effective.}
    \label{fig: test_recoveredVSgruner}
\end{figure}

Despite all 89 blended pairs being simulated from real WBs known to follow gyrochronology, only 34 pairs (38.2\%) satisfied the CC. This discrepancy highlights the challenges introduced by signal blending, noise, and peak selection. Dominant peaks in the periodogram may not correspond to the true periods of either component, and the correct periods may be present but not selected due to their lower amplitudes or proximity to other peaks. Additionally, the injected Gaussian noise and smoothing may attenuate or shift the signals enough to move stars away from the expected rotation-age relation. These results illustrate the limitations of recovering accurate rotation periods from blended systems, even when the underlying physics is consistent with gyrochronology.

Focusing on the 34 coeval pairs (68 individual stars), Figure \ref{fig: test_recoveredVSgruner} shows that our CC method successfully recovers most rotation periods (64.7\%) from Gruner's Kepler/K2 WB sample, especially in the region where TESS is most sensitive ($P_{rot} < 12$ days; shaded area). Within this region, our method shows its strongest performance, correctly recovering 88\% of the rotation periods (21 out of 24 stars). Across the entire plot, when considering both components as a pair instead of single stars, the recovery decreases to 44.1\%. This highlights that accurate recovery for at least one component remains likely, even in blended scenarios.

\begin{deluxetable}{cccc}
\tablecaption{Frequency and accuracy of selected peak combinations in the simulated blend test. Each combination refers to the selected Lomb-Scargle (LS) peaks for the primary and secondary components, respectively (e.g., LS1LS2 indicates the first peak for the primary and the second peak assigned to the secondary). ``Selected" indicates how often each combination was chosen by the method, ``Matched" refers to cases where the selected periods match the true ones from the original (unblended) data, and ``\% Correct" gives the success rate for each combination. \label{tab: combination_statistics}}

\tablehead{
\colhead{Combination} & \colhead{Selected} & \colhead{Matched} & \colhead{\% Correct}
}
\startdata
LS1LS1 & 0 & 0 & 0.00 \\
LS1LS2 & 13 & 8 & 61.54 \\
LS1LS3 & 4 & 2 & 50.00 \\
LS2LS1 & 4 & 1 & 25.00 \\
LS2LS2 & 0 & 0 & 0.00 \\
LS2LS3 & 6 & 2 & 33.33 \\
LS3LS1 & 6 & 2 & 33.33 \\
LS3LS2 & 1 & 0 & 0.00 \\
LS3LS3 & 0 & 0 & 0.00 \\
\hline
Total  & 34 & 15 & 38.2 \\
\enddata
\end{deluxetable}

To better understand which combinations of periodogram peaks lead to correct period recovery, we examined the frequency of each selected pair among the first three LS peaks of each WB component. In the 34 blends that satisfy our coevality criterion, 15 are unambiguous—only a single LS-peak pairing passes the test—and in 4 of those, the solution is LS1LS2 (matching Gruner’s benchmark in 2 cases, or 50\%). In the remaining 19 ambiguous blends, we applied the tie-breaker rule. This produced a mix of pairings: nine LS1LS2, four LS1LS3, three LS2LS3, two LS3LS1, and one LS2LS1. While this tie-breaker does introduce a mild bias toward the brighter star’s signal, it increased our usable sample by 21\% and recovered the true period in 17 of these difficult cases. This outcome demonstrates that our selection rule strikes a practical balance between completeness and reliability.

Table~\ref{tab: combination_statistics} summarizes the occurrence and success rate for each combination. The most frequent and most successful combination is LS1LS2, which was selected 13 times and yielded the correct period in 61.5\% of cases. This trend indicates that the first and second peaks (LS1 and LS2) are more likely to include the true rotation periods. Although the prevalence of LS1LS2 is partly driven by our tie-breaking rule, its ability to recover the true period in over 60\% of ambiguous blends demonstrates its validity as a primary estimator under ambiguity.

LS1LS3 also shows a moderate success rate (50.0\%), although it is less frequently selected. Combinations involving LS2LS2 and LS3LS3 were never selected, but this is expected, as these combinations reflect blended scenarios (horizontal lines in the CPD) and such combinations usually disagree with the CC. 

\begin{figure}[ht!]
    \centering
    \includegraphics[width=0.9\linewidth]{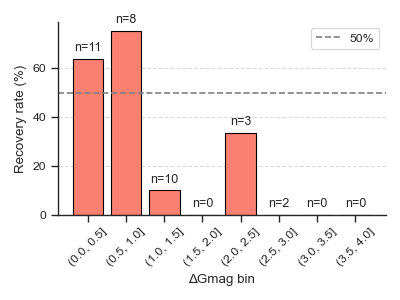}
    \caption{Recovery rate as a function of $\Delta{Gmag}$. Bins with recovery = 0 still contain systems, highlighting that correct period assignment fails for high $\Delta{Gmag}$ ranges.}
    \label{fig: test_deltam}
\end{figure}

Although combinations such as LS2LS1, LS2LS3, and LS3LS1 show relatively low success rates, they are particularly informative. In cases like LS2LS1 and LS3LS1, the first period corresponds to the fainter star dominating the light curve, suggesting scenarios where increased spot coverage on the dimmer component makes its signal more prominent despite its lower flux contribution. These cases highlight that the brightest star does not always dominate the observed variability and that physical factors like spot distribution can shift the balance of detectability. The CC helps uncover these subtle cases, as it allows us to recognize when the recovered periods are consistent with a coeval age despite non-intuitive flux dominance.

We also examined how the brightness contrast between components ($\Delta{Gmag}$) affects period recovery (See Figure \ref{fig: test_deltam}). Systems with $\Delta{Gmag}< 1.0$  had the highest recovery rates (64–75\%), while recovery dropped sharply for larger magnitude differences. This trend illustrates the expectation that flux contrast is a limiting factor: as one star increasingly dominates the blended light curve, the fainter component’s signal is more likely to be suppressed or misidentified. 

\section{Period Recovery in a Contaminated World: Insights from Our TESS Sample} \label{sec: results}

\begin{figure}[h]
    \centering
    \includegraphics[width=\linewidth]{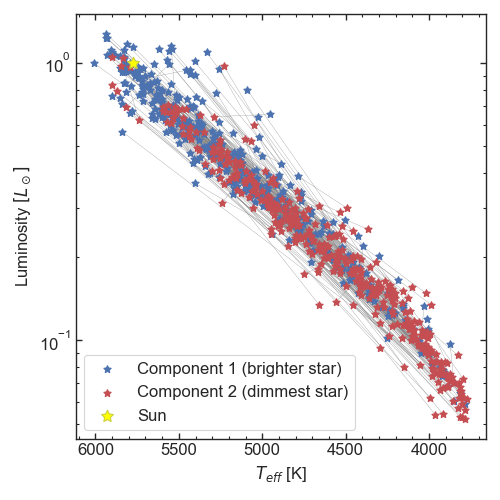}
    \caption{HR diagram of our 360 MS+MS WB pairs sample. Components of each pair are connected by lines.}
    \label{fig: sample}
\end{figure}

\begin{figure*}[t!]
    \centering
    \includegraphics[width=\textwidth]{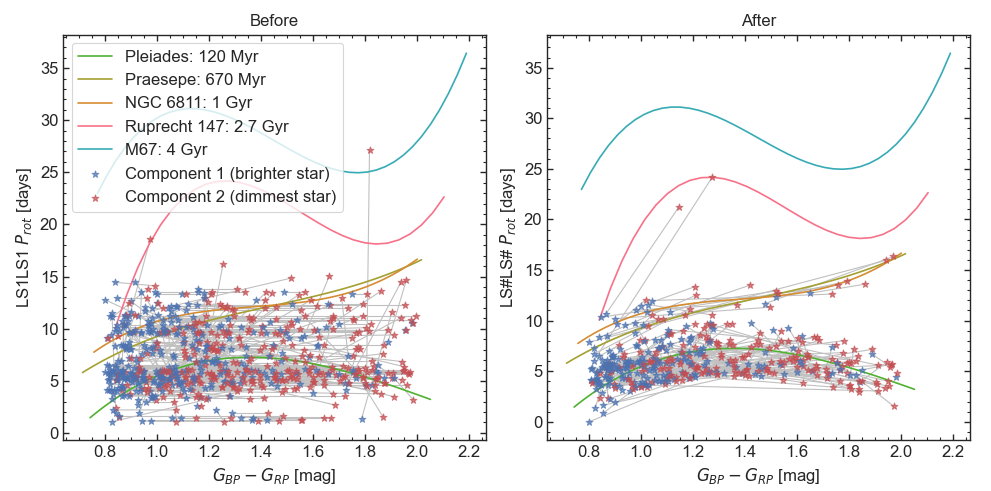}
    \caption{CPD before and after applying the CC method.}
    \label{fig: before and after}
\end{figure*}

In this section, we evaluate the performance of the CC applied to real WBs observed by TESS. Figure \ref{fig: sample} shows the Hertzsprung–Russell (HR) diagram for the full sample of the WBs analyzed. It comprises 360 MS+MS WB pairs drawn from every TESS cycle available up to cycle 3. These pairs have Gaia color indexes $\in [0.75–2.0]$ ($T_{eff}\sim[3400-6200]$ K), between the fully convective boundary and the Kraft break, where the gyrochronology paradigm applies. All 360 WB pairs are in the \cite{El-Badry2021} catalog, showing $R<0.1$, meaning a high degree of confidence of being physical pairs. 

We constructed our light curves by creating custom masks for each binary component using the approach outlined in \cite{Nielsen2020} and \cite{Metcalfe2023}. In this approach,  the brightest pixel is assigned to each target, and the mask is expanded one pixel at a time until the noise level in the extracted light curve reaches its minimum value. We then addressed spacecraft systematics using an iterative detrending approach based on the correlation of the flux with the image centroid location. To prepare the light curves for rotation period analysis, we normalized each sector relative to its median flux and filled the gaps using spline interpolation. For all the light curves generated, we prewhitened the first three peaks with the highest amplitudes. In addition, we extracted false periods known to affect TESS time series data \citep{Lares-Martiz2024} to isolate the effects of blending. 

Only 52 of 360 pairs exhibited initial agreement with gyrochronology expectations, meaning that selecting the dominant peak (LS1) from each component's periodogram as rotation periods, the pair matched a common gyrochronology isochrone within $2\sigma$. The rest of the pairs (308) violate our CC, likely due to unresolved blends (221 pairs showed nearly horizontal lines in the CPD, see left plot in Figure \ref{fig: before and after}), spurious period detections, hierarchical systems, or because they don't have the ages of the open clusters used as references. After applying the period reconciliation step (right plot in Figure \ref{fig: before and after}), this number increased to 269 pairs, demonstrating the ability of the approach to recover plausible rotation periods even in cases with initial mismatches due to contamination, blending, or period misidentification.

\subsection{Kinematical constraints} \label{subsec:kinematics}

\begin{figure}[t!]
    \centering
    \includegraphics[scale=0.38]{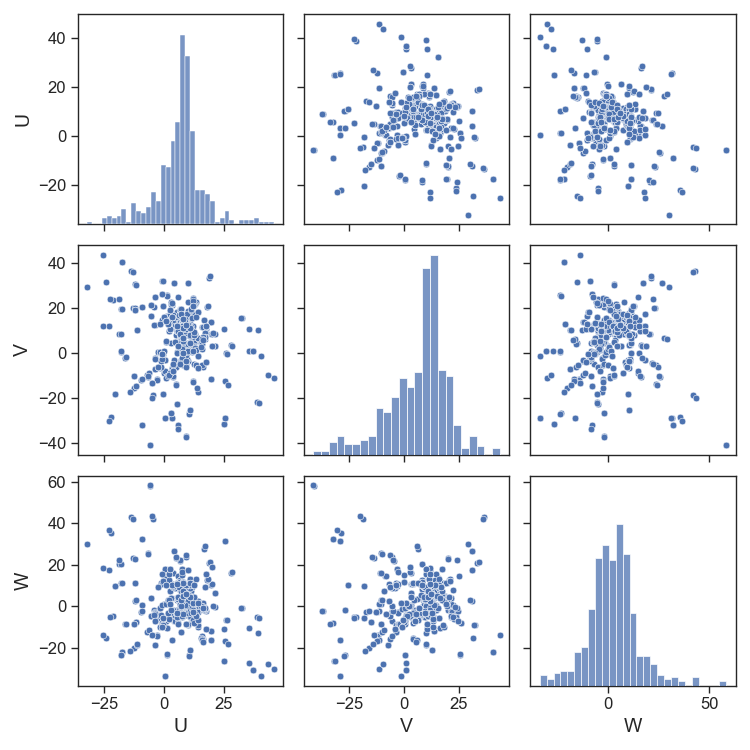}
    \caption{Distribution of Galactic space velocities (U, V, W) in km/s for the same stars.}
    \label{fig: kinematic_ages}
\end{figure}

Since stellar velocity dispersions increase with age due to kinematical interactions in the Galaxy, these motions can be used to infer stellar ages statistically. Figure \ref{fig: kinematic_ages} shows the distribution of our WB sample galactic velocity components, using Gaia astrometry for the initially agreeing and reconciled pairs. The histograms reveal that the distributions of U and W are symmetric and centered near $0$ km/s, as expected for a kinematically young population ($\leq2$ Gyr). They are tightly clustered near the origin of the plot, with most having $|W| < 20$ km/s, indicating they belong to a kinematically cold population typical of the thin disk \citep{Bensby2003, Nordstrom2004}. The V velocity, which traces motion in the direction of Galactic rotation, exhibits a peak slightly skewed toward positive values. This behavior is consistent with expectations for young ($\leq2$ Gyr) thin disk stars, which typically have small velocity dispersions and exhibit minimal asymmetric drift. In fact, studies have shown that stars younger than $\sim1$ Gyr often have V velocities close to or slightly exceeding the Local Standard of Rest \citep{Dehnen1998,Nordstrom2004,Casagrande2011}. 

While \cite{Hattori2025} has demonstrated that TESS can recover rotation periods longer than 10 days, these cases rely on nearly continuous TESS coverage and external ground-based confirmation. In general, however, TESS is most reliable for detecting short periods ($\lesssim10-13$ days), as shown by \citet{Avallone2022} and \citet{Boyle2025}. Thus, our sample is inherently biased toward stars younger than $\sim1$ Gyr; the fact that these targets also exhibit thin‐disk kinematics ($|W| < 20$ km/s) both confirms this expectation and supports the validity of our period determinations and coevality criterion, even in the presence of moderate contamination.

\begin{figure}[h]
    \centering
    \includegraphics[width=1.0\linewidth]{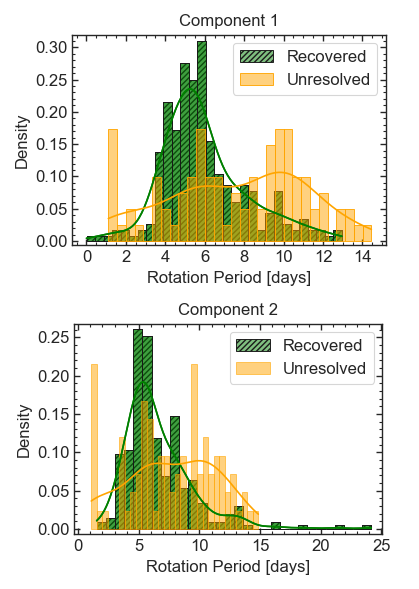}
    \caption{Histogram of rotation periods per component of the recovered and unresolved pairs}
    \label{fig: prot_distribution}
\end{figure}

Figure \ref{fig: prot_distribution} illustrates that for both components, the recovered rotation periods predominantly cluster between approximately 3 and 8 days \citep[also noted by][]{Oswalt2022}. Longer rotation periods become increasingly challenging to recover reliably, as evidenced by the broader and more uniform distribution of unresolved cases extending beyond 8–10 days. Specifically, Component 1 shows almost no successful recoveries beyond $\sim10$ days, while Component 2 similarly experiences a sharp decline in recoverability beyond roughly the same limit. This limit to TESS period detection was first noted by \cite{Avallone2022}

Recent Galactic‐scale studies offer another explanation for the observed excess near 5 days. The white-dwarf formation rate has accelerated by a factor of 2–3 over the last 1–2 Gyr, boosting the number of WD–MS wide binaries whose MS companions share that age \citep{Holberg2016}. In addition, a distinct burst of star formation $\sim1$ Gyr ago was identified \citep{Chen2025,Fantin2019}. Stars of that epoch should today spin with periods near $\sim5$ days. Together, these effects imply that our TESS WB sample may be especially rich in genuine rotation signals around 5 days, while older bursts correspond to longer periods that lie beyond TESS’s optimal detection window.

These findings therefore indicate a practical threshold for rotation-period recovery from TESS blended observations: periods shorter than approximately 8 days are reliably detectable, whereas periods longer than about 10 days are significantly more challenging and prone to remaining unresolved. This limitation is consistent with the sector duration of TESS observations ($\sim27$ days), as longer periods produce fewer cycles within a given window, diminishing detectability in blended scenarios. Recent studies corroborate this threshold: \cite{Avallone2022} observed that rotation-period reliability falls off sharply beyond $\sim13.7,$ days, coinciding with the TESS orbital period, while \cite{Boyle2025} demonstrated a similar decline beyond $\sim10\mbox{–}12,$ days due to sector-duration constraints. Additionally, the presence of unresolved pairs (for which the rotation periods cannot be confidently recovered or distinguished) at shorter periods ($<5$ days) underscores the importance of contamination levels and the intrinsic amplitude variability of stellar signals, which can fluctuate due to changes in star spot coverage or activity levels.

\begin{figure}[h!]
    \centering
    \includegraphics[scale=1]{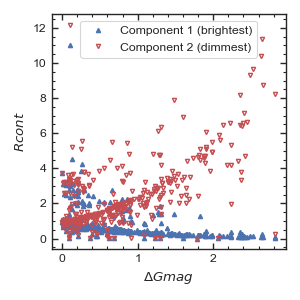}
    \caption{Contamination ratio (Rcont) as a function of the magnitude difference ($\Delta G{mag}$) between the components of each binary.}
    \label{fig: contaminationVSDeltam}
\end{figure}

To further assess the impact of contamination, Figure~\ref{fig: contaminationVSDeltam} shows the magnitude difference between the two components of each binary ($\Delta G{\mathrm{mag}}$) as a function of the contamination ratio (Rcont), defined as the total contaminant flux divided by the target star flux \citep{Stassun2018}. As expected, the dimmer component exhibits a higher Rcont, suggesting that the rotation signal of the brighter star is more likely to be recovered reliably, given its typically lower contamination levels.

\begin{figure}[h]
    \centering
    \includegraphics[width=1.0\linewidth]{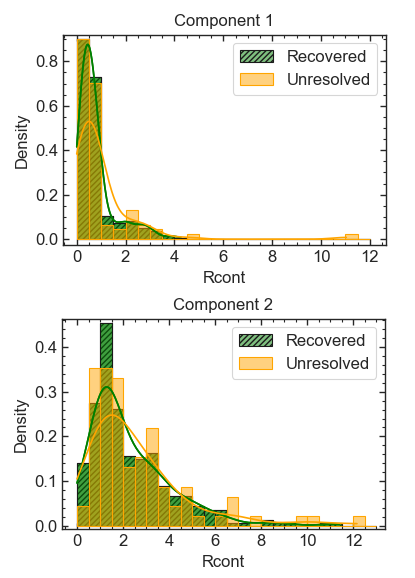}
    \caption{Histogram of Rcont per component of the recovered and unresolved pairs}
    \label{fig: contamination}
\end{figure}

The critical role of the brighter star's contamination level for successful rotation period recovery in TESS blended observations is highlighted in Figure \ref{fig: contamination}. Successful period recovery mainly depends on the contamination of the brighter component (Component 1), sharply declining when its contamination exceeds unity. In contrast, contamination of the fainter component (Component 2) shows a broader distribution, indicating, as expected, that its period is more difficult to recover.

\begin{figure}[h]
   \centering
    \includegraphics[width=1.0\linewidth]{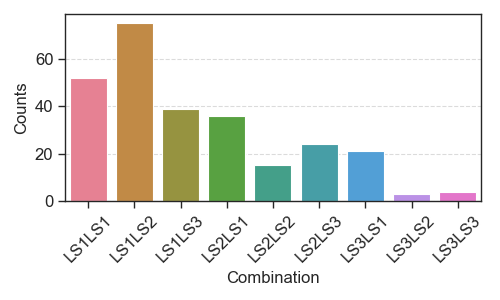}
   \caption{Occurrences of combinations}
    \label{fig: ocurrences}
\end{figure}

Finally, we investigated the frequency of different combinations of periodogram peaks selected for each successful period recovery using CC (Figure \ref{fig: ocurrences}). In practice, the LS1LS2 pairing is the most statistically robust choice for coevality enforcement, as the highest‐amplitude peak (LS1) typically corresponds to the true dominant rotation signal, and the second‐highest peak (LS2) captures the companion’s modulation. Nonetheless, the non-negligible occurrence of LS2LS1 and LS3LS1 combinations—where the fainter star’s peak exceeds that of its brighter companion—underscores that spot‐coverage, geometry, or inclination can occasionally elevate the dimmer star’s signal. 

\section{Summary and Implications for Stellar Rotation Studies} \label{sec: discussion}

In this work, we evaluated the reliability of stellar rotation periods derived from TESS light curves in WBs, using gyrochronology as a physical constraint. We assumed that the components of each WB are coeval and therefore that their rotation periods should be consistent with a single gyrochrone. This allowed us to assess the robustness of period detections, particularly in cases where light curves are blended.

In our simulated blends based on Kepler light curves, we found that our method achieves an 88\% success rate in recovering correct stellar rotation periods within the regime where TESS detections are most reliable ($P_\mathrm{rot} < 12$ days). We found that certain combinations of periodogram peaks yield rotation periods that align within $2\sigma$ of a common gyrochrone, indicating that rotational signals can, in some cases, be successfully disentangled despite significant blending. The highest recovery rates typically occur when assigning the dominant periodogram peak to the brighter star and the secondary peak to the fainter component (i.e. LS1LS2), consistent with the expectation that variability amplitude scales with flux contribution. However, significant match rates also occur for combinations where the fainter star dominates the signal, likely due to higher spot coverage or a more favorable inclination enhancing its rotational modulation. 

Importantly, our analysis provides a practical threshold for rotation-period recovery from TESS blended observations: periods shorter than approximately 8 days can be reliably detectable, whereas periods longer than about 10 days are significantly more challenging and prone to remaining unresolved. This limitation is consistent with the sector duration of TESS observations (~27 days), as longer periods produce fewer cycles within the available time baseline.

Our investigation has significant implications for detecting rotational modulation in TESS light curves. Many automated pipelines assume that the dominant periodogram peak is the rotation period of the primary target. We have shown that this assumption often fails, even when contamination levels are modest but non-negligible. Therefore, caution is required when interpreting periodograms derived from light curves extracted from stars in crowded fields such as clusters or WBs. The coevality constraint (CC) used here is a physically motivated framework for assigning the most likely rotation periods. While it does not guarantee a correct solution in every case, it enables more informed period selections and provides a means of avoiding many false positives.

Gyrochronology constraints can help evaluate the plausibility of different period combinations within a physically motivated framework, but their effectiveness is limited by contamination and the complexity of the periodogram structure. Although rotational periods can sometimes be recovered despite blending, our results exhibit sensitivity to parameter choices, including the selection of candidate peaks and tolerances in the open cluster fits.

Gyrochronology, when applied in this novel way, can play a critical role in guiding period validation, offering a physically motivated constraint that becomes increasingly valuable in the blended, short time-baseline, high-volume regime of TESS data. In our study, the scatter observed in the CC corrected gyrochrones (Figure \ref{fig: before and after}, “After”) is comparable to that seen in open clusters in the CPD. While this similarity reflects the selection imposed by our method, it demonstrates that, for systems meeting the CC, gyrochronology has the potential to yield age estimates with precision comparable to cluster-based approaches. While the specific focus of this work is on wide binaries, the conclusions are equally relevant for single field stars, where contamination and blending may also complicate period recovery. Additionally, applying gyrochronology in this way helps mitigate the impact of instrumental false periods introduced by TESS systematics. Gyrochronology may not provide all the answers, but in the age of TESS, it remains one of the most powerful tools we have for interpreting stellar rotation data.

\begin{acknowledgments}
We thank Zach Claytor and David Gruner for their support. Also, special thanks to Sofía Rocha and Diego Garrido.  Support from NSF grants AST-1910396, AST-2108975 and NASA grants 80NSSC22K0622, 80NSSC21K0245, and NNX16AB76G is gratefully acknowledged.

\end{acknowledgments}

\bibliography{mlm_erau2025}{}

\begin{thebibliography}{}
\expandafter\ifx\csname natexlab\endcsname\relax\def\natexlab#1{#1}\fi
\providecommand{\url}[1]{\href{#1}{#1}}
\providecommand{\dodoi}[1]{doi:~\href{http://doi.org/#1}{\nolinkurl{#1}}}
\providecommand{\doeprint}[1]{\href{http://ascl.net/#1}{\nolinkurl{http://ascl.net/#1}}}
\providecommand{\doarXiv}[1]{\href{https://arxiv.org/abs/#1}{\nolinkurl{https://arxiv.org/abs/#1}}}

\bibitem[{E.~A. {Avallone} {et~al.}(2022){Avallone}, {Tayar}, {van Saders}, {Berger}, {Claytor}, {Beaton}, {Teske}, {Godoy-Rivera}, \& {Pan}}]{Avallone2022}
{Avallone}, E.~A., {Tayar}, J.~N., {van Saders}, J.~L., {et~al.} 2022, \bibinfo{title}{{Rotation Distributions around the Kraft Break with TESS and Kepler: The Influences of Age, Metallicity, and Binarity},} \apj, 930, 7, \dodoi{10.3847/1538-4357/ac60a1}

\bibitem[{S.~A. {Barnes}(2003){Barnes}}]{Barnes2003}
{Barnes}, S.~A. 2003, \bibinfo{title}{{On the Rotational Evolution of Solar- and Late-Type Stars, Its Magnetic Origins, and the Possibility of Stellar Gyrochronology},} \apj, 586, 464, \dodoi{10.1086/367639}

\bibitem[{T. {Bensby} {et~al.}(2003){Bensby}, {Feltzing}, \& {Lundstr{\"o}m}}]{Bensby2003}
{Bensby}, T., {Feltzing}, S., \& {Lundstr{\"o}m}, I. 2003, \bibinfo{title}{{Elemental abundance trends in the Galactic thin and thick disks as traced by nearby F and G dwarf stars},} \aap, 410, 527, \dodoi{10.1051/0004-6361:20031213}

\bibitem[{A.~S. {Binks} \& H.~M. {G{\"u}nther}(2024){Binks} \& {G{\"u}nther}}]{Binks2024}
{Binks}, A.~S., \& {G{\"u}nther}, H.~M. 2024, \bibinfo{title}{{TESSILATOR: a one-stop shop for measuring TESS rotation periods},} \mnras, 533, 2162, \dodoi{10.1093/mnras/stae1850}

\bibitem[{L.~G. Bouma {et~al.}(2023)Bouma, Palumbo, \& Hillenbrand}]{Bouma2023}
Bouma, L.~G., Palumbo, E.~K., \& Hillenbrand, L.~A. 2023, \bibinfo{title}{The Empirical Limits of Gyrochronology,} The Astrophysical Journal Letters, 947, L3, \dodoi{10.3847/2041-8213/acc589}

\bibitem[{A.~W. {Boyle} {et~al.}(2025){Boyle}, {Mann}, \& {Bush}}]{Boyle2025}
{Boyle}, A.~W., {Mann}, A.~W., \& {Bush}, J. 2025, \bibinfo{title}{{Quantifying the Limits of TESS Stellar Rotation Measurements with the K2-TESS Overlap},} arXiv e-prints, arXiv:2504.13262, \dodoi{10.48550/arXiv.2504.13262}

\bibitem[{A.~G.~A. Brown {et~al.}(2021)Brown, Vallenari, Prusti, Bruijne, Babusiaux, Biermann, Creevey, Evans, Eyer, Hutton, Jansen, Jordi, Klioner, Lammers, Lindegren, Luri, Mignard, Panem, Pourbaix, Randich, Sartoretti, Soubiran, Walton, Arenou, Bailer‐Jones, Bastian, Cropper, Drimmel, Katz, Lattanzi, Leeuwen, Bakker, Cacciari, Castañeda, Angeli, Ducourant, Fabricius, Fouesneau, Frémat, Guerra, Guerrier, Guiraud, Piccolo, Masana, Messineo, Mowlavï, Nicolas, Nienartowicz, Pailler, Panuzzo, Riclet, Roux, Seabroke, Sordo, Tanga, Thévenin, Gracia-Abril, Portell, Teyssier, Altmann, Andrae, Bellas‐Velidis, Benson, Berthier, Blomme, Brugaletta, Burgess, Busso, Carry, Cellino, Cheek, Clementini, Damerdji, Davidson, Delchambre, Dell’Oro, Fernández-Hernández, Galluccio, García-Lario, Garcia-Reinaldos, González-Núñez, Gosset, Haigron, Halbwachs, Hambly, Harrison, Hatzidimitriou, Heiter, Hernández, Hestroffer, Hodgkin, Holl, Janßen, Fombelle, Jordan, Krone-Martins, Lanzafame, Löffler, Lorca,
  Manteiga, Marchal, Marrese, Moitinho, Mora, Muinonen, Osborne, Pancino, Pauwels, Petit, Recio–Blanco, Richards, Riello, Rimoldini, Robin, Roegiers, Rybizki, Sarro, Siopis, Smith, Sozzetti, Ulla, Utrilla, Leeuwen, Reeven, Abbas, Aramburu, Accart, Aerts, Aguado, Ajaj, Altavilla, Álvarez, Cid-Fuentes, Alves, Anderson, Varela, Antoja, Audard, Baines, Baker, Balaguer-Núñez, Balbinot, Balog, Barache, Barbato, Barros, Barstow, Bartolomé, Bassilana, Bauchet, Baudesson-Stella, Becciani, Bellazzini, Bernet, Bertone, Bianchi, Blanco-Cuaresma, Boch, Bombrun, Bossini, Bouquillon, Bragaglia, Bramante, Breedt, Bressan, Brouillet, Bucciarelli, Burlacu, Busonero, Butkevich, Buzzi, Caffau, Cancelliere, Cánovas, Cantat-Gaudin, Carballo, Carlucci, Carnerero, Carrasco, Casamiquela, Castellani, Castro-Ginard, Sampol, Chaoul, Charlot, Chemin, Chiavassa, Cioni, Comoretto, Cooper, Cornez, Cowell, Crifo, Crosta, Crowley, Dafonte, Dapergolas, David, David, Laverny, Luise, March, Ridder, Souza, Teodoro, Torres, Peloso, Pozo,
  Delbò, Delgado, Delgado, Delisle, Matteo, Diakité, Diener, Distefano, Dolding, Eappachen, Edvardsson, Enke, Esquej, Fabre, Fabrizio, Faigler, Fedorets, Fernique, Fienga, Figueras, Fouron, Fragkoudi, Fraile, Franke, Gai, Garabato, Garcia-Gutierrez, García-Torres, Garofalo, Gavras, Gerlach, Geyer, Giacobbe, Gilmore, Girona, Giuffrida, Gomel, Gómez, González-Santamaría, González–Vidal, Granvik, Gutiérrez–Sánchez, Guy, Hauser, Haywood, Helmi, Hidalgo, Hilger, Hładczuk, Hobbs, Holland, Huckle, Jasniewicz, Jonker, Campillo, Julbé, Karbevska, Kervella, Khanna, Kochoska, Kontizas, Kordopatis, Korn, Kostrzewa-Rutkowska, Kruszyńska, Lambert, Lanza, Lasne, Campion, Fustec, Lebreton, Lebzelter, Leccia, Leclerc, Lecœur-Taı̈bi, Liao, Licata, Lindstrøm, Lister, Livanou, Lobel, Pardo, Managau, Mann, Marchant, Marconi, Santos, Marinoni, Marocco, Marshall, Polo, Martín–Fleitas, Masip, Massari, Mastrobuono-Battisti, Mazeh, McMillan, Messina, Michalik, Millar, Mints, Molina, Molinaro, Molnár,
  Montegriffo, Mor, Morbidelli, Morel, Morris, Mulone, Muñoz, Muraveva, Murphy, Musella, Noval, Ordénovic, Orrù, Osinde, Pagani, Pagano, Palaversa, Palicio, Panahi, Pawlak, Esteller, Penttilä, Piersimoni, Pineau, Plachy, Plum, Poggio, Poretti, Poujoulet, Prša, Pulone, Racero, Ragaini, Rainer, Raiteri, Rambaux, Ramos, Ramos-Lerate, Fiorentin, Regibo, Reylé, Ripepi, Riva, Rixon, Robichon, Ciardullo, Roelens, Rohrbasser, Romero-Gómez, Rowell, Royer, Rybicki, Sadowski, Sellés, Sahlmann, Salgado, Salguero, Samaras, Gimenez, Sanna, Santoveña, Sarasso, Schultheis, Sciacca, Segol, Segovia, Ségransan, Semeux, Shahaf, Siddiqui, Siebert, Siltala, Slezak, Smart, Solano, Solitro, Souami, Souchay, Spagna, Spoto, Steele, Steidelmüller, Stephenson, Süveges, Szabados, Szegedi-Elek, Taris, Tauran, Taylor, Teixeira, Thuillot, Tonello, Torra, Torra, Turon, Unger, Vaillant, Dillen, Vanel, Vecchiato, Viala, Vicente, Voutsinas, Weiler, Wevers, Wyrzykowski, Yoldaş, Yvard, Zhao, Zorec, Zucker, Zurbach, \&
  Zwitter}]{Brown2021}
Brown, A. G.~A., Vallenari, A., Prusti, T., {et~al.} 2021, \bibinfo{title}{<i>Gaia</I> Early Data Release 3,} Astronomy and Astrophysics, \dodoi{10.1051/0004-6361/202039657e}

\bibitem[{L. {Casagrande} {et~al.}(2011){Casagrande}, {Sch{\"o}nrich}, {Asplund}, {Cassisi}, {Ram{\'\i}rez}, {Mel{\'e}ndez}, {Bensby}, \& {Feltzing}}]{Casagrande2011}
{Casagrande}, L., {Sch{\"o}nrich}, R., {Asplund}, M., {et~al.} 2011, \bibinfo{title}{{New constraints on the chemical evolution of the solar neighbourhood and Galactic disc(s). Improved astrophysical parameters for the Geneva-Copenhagen Survey},} \aap, 530, A138, \dodoi{10.1051/0004-6361/201016276}

\bibitem[{T. {Chen} \& N. {Prantzos}(2025){Chen} \& {Prantzos}}]{Chen2025}
{Chen}, T., \& {Prantzos}, N. 2025, \bibinfo{title}{{Recent star formation episodes in the Galaxy: Impact on its chemical properties and the evolution of its abundance gradient},} \aap, 694, A120, \dodoi{10.1051/0004-6361/202452552}

\bibitem[{Z.~R. {Claytor} {et~al.}(2022){Claytor}, {van Saders}, {Llama}, {Sadowski}, {Quach}, \& {Avallone}}]{Claytor2022}
{Claytor}, Z.~R., {van Saders}, J.~L., {Llama}, J., {et~al.} 2022, \bibinfo{title}{{Recovery of TESS Stellar Rotation Periods Using Deep Learning},} \apj, 927, 219, \dodoi{10.3847/1538-4357/ac498f}

\bibitem[{J.~L. {Curtis} {et~al.}(2019){Curtis}, {Ag{\"u}eros}, {Douglas}, \& {Meibom}}]{Curtis2019}
{Curtis}, J.~L., {Ag{\"u}eros}, M.~A., {Douglas}, S.~T., \& {Meibom}, S. 2019, \bibinfo{title}{{A Temporary Epoch of Stalled Spin-down for Low-mass Stars: Insights from NGC 6811 with Gaia and Kepler},} \apj, 879, 49, \dodoi{10.3847/1538-4357/ab2393}

\bibitem[{J.~L. {Curtis} {et~al.}(2020){Curtis}, {Ag{\"u}eros}, {Matt}, {Covey}, {Douglas}, {Angus}, {Saar}, {Cody}, {Vanderburg}, {Law}, {Kraus}, {Latham}, {Baranec}, {Riddle}, {Ziegler}, {Lund}, {Torres}, {Meibom}, {Aguirre}, \& {Wright}}]{Curtis2020}
{Curtis}, J.~L., {Ag{\"u}eros}, M.~A., {Matt}, S.~P., {et~al.} 2020, \bibinfo{title}{{When Do Stalled Stars Resume Spinning Down? Advancing Gyrochronology with Ruprecht 147},} \apj, 904, 140, \dodoi{10.3847/1538-4357/abbf58}

\bibitem[{W. {Dehnen} \& J.~J. {Binney}(1998){Dehnen} \& {Binney}}]{Dehnen1998}
{Dehnen}, W., \& {Binney}, J.~J. 1998, \bibinfo{title}{{Local stellar kinematics from HIPPARCOS data},} \mnras, 298, 387, \dodoi{10.1046/j.1365-8711.1998.01600.x}

\bibitem[{S.~T. {Douglas} {et~al.}(2017){Douglas}, {Ag{\"u}eros}, {Covey}, \& {Kraus}}]{Douglas2017}
{Douglas}, S.~T., {Ag{\"u}eros}, M.~A., {Covey}, K.~R., \& {Kraus}, A. 2017, \bibinfo{title}{{Poking the Beehive from Space: K2 Rotation Periods for Praesepe},} \apj, 842, 83, \dodoi{10.3847/1538-4357/aa6e52}

\bibitem[{S.~T. {Douglas} {et~al.}(2019){Douglas}, {Curtis}, {Ag{\"u}eros}, {Cargile}, {Brewer}, {Meibom}, \& {Jansen}}]{Douglas2019}
{Douglas}, S.~T., {Curtis}, J.~L., {Ag{\"u}eros}, M.~A., {et~al.} 2019, \bibinfo{title}{{K2 Rotation Periods for Low-mass Hyads and a Quantitative Comparison of the Distribution of Slow Rotators in the Hyades and Praesepe},} \apj, 879, 100, \dodoi{10.3847/1538-4357/ab2468}

\bibitem[{K. {El-Badry} {et~al.}(2021){El-Badry}, {Rix}, \& {Heintz}}]{El-Badry2021}
{El-Badry}, K., {Rix}, H.-W., \& {Heintz}, T.~M. 2021, \bibinfo{title}{{A million binaries from Gaia eDR3: sample selection and validation of Gaia parallax uncertainties},} \mnras, 506, 2269, \dodoi{10.1093/mnras/stab323}

\bibitem[{N.~J. {Fantin} {et~al.}(2019){Fantin}, {C{\^o}t{\'e}}, {McConnachie}, {Bergeron}, {Cuillandre}, {Gwyn}, {Ibata}, {Thomas}, {Carlberg}, {Fabbro}, {Haywood}, {Lan{\c{c}}on}, {Lewis}, {Malhan}, {Martin}, {Navarro}, {Scott}, \& {Starkenburg}}]{Fantin2019}
{Fantin}, N.~J., {C{\^o}t{\'e}}, P., {McConnachie}, A.~W., {et~al.} 2019, \bibinfo{title}{{The Canada-France Imaging Survey: Reconstructing the Milky Way Star Formation History from Its White Dwarf Population},} \apj, 887, 148, \dodoi{10.3847/1538-4357/ab5521}

\bibitem[{D. {Gruner} {et~al.}(2023{\natexlab{a}}){Gruner}, {Barnes}, \& {Janes}}]{Gruner2023}
{Gruner}, D., {Barnes}, S.~A., \& {Janes}, K.~A. 2023{\natexlab{a}}, \bibinfo{title}{{Wide binaries demonstrate the consistency of rotational evolution between open cluster and field stars},} \aap, 675, A180, \dodoi{10.1051/0004-6361/202346590}

\bibitem[{D. {Gruner} {et~al.}(2023{\natexlab{b}}){Gruner}, {Barnes}, \& {Weingrill}}]{Gruner2023M67}
{Gruner}, D., {Barnes}, S.~A., \& {Weingrill}, J. 2023{\natexlab{b}}, \bibinfo{title}{{New insights into the rotational evolution of near-solar age stars from the open cluster M 67},} \aap, 672, A159, \dodoi{10.1051/0004-6361/202345942}

\bibitem[{T. {Han} \& T.~D. {Brandt}(2023){Han} \& {Brandt}}]{Han2023}
{Han}, T., \& {Brandt}, T.~D. 2023, \bibinfo{title}{{TESS-Gaia Light Curve: A PSF-based TESS FFI Light-curve Product},} \aj, 165, 71, \dodoi{10.3847/1538-3881/acaaa7}

\bibitem[{S. {Hattori} {et~al.}(2025){Hattori}, {Angus}, {Foreman-Mackey}, {Yuxi}, {Lu}, \& {Colman}}]{Hattori2025}
{Hattori}, S., {Angus}, R., {Foreman-Mackey}, D., {et~al.} 2025, \bibinfo{title}{{Measuring Long Stellar Rotation Periods (>10 days) from TESS FFI Light Curves is Possible: An Investigation Using TESS and ZTF},} arXiv e-prints, arXiv:2505.10376, \dodoi{10.48550/arXiv.2505.10376}

\bibitem[{C. Hedges {et~al.}(2020)Hedges, Angus, Barentsen, Saunders, Montet, \& Gully-Santiago}]{Hedges2020}
Hedges, C., Angus, R., Barentsen, G., {et~al.} 2020, \bibinfo{title}{Systematics-insensitive Periodogram for Finding Periods in TESS Observations of Long-period Rotators,} Research Notes of the AAS, 4, 220, \dodoi{10.3847/2515-5172/abd106}

\bibitem[{C. {Hedges} {et~al.}(2021){Hedges}, {Luger}, {Martinez-Palomera}, {Dotson}, \& {Barentsen}}]{Hedges2021}
{Hedges}, C., {Luger}, R., {Martinez-Palomera}, J., {Dotson}, J., \& {Barentsen}, G. 2021, \bibinfo{title}{{Linearized Field Deblending: Point-spread Function Photometry for Impatient Astronomers},} \aj, 162, 107, \dodoi{10.3847/1538-3881/ac0825}

\bibitem[{M.~E. {Higgins} \& K.~J. {Bell}(2023){Higgins} \& {Bell}}]{Higgins2023}
{Higgins}, M.~E., \& {Bell}, K.~J. 2023, \bibinfo{title}{{Localizing Sources of Variability in Crowded TESS Photometry},} \aj, 165, 141, \dodoi{10.3847/1538-3881/acb20c}

\bibitem[{J.~B. {Holberg} {et~al.}(2016){Holberg}, {Oswalt}, {Sion}, \& {McCook}}]{Holberg2016}
{Holberg}, J.~B., {Oswalt}, T.~D., {Sion}, E.~M., \& {McCook}, G.~P. 2016, \bibinfo{title}{{The 25 parsec local white dwarf population},} \mnras, 462, 2295, \dodoi{10.1093/mnras/stw1357}

\bibitem[{M. {Lares-Martiz} {et~al.}(2024){Lares-Martiz}, {Buzasi}, {Oswalt}, {Confeiteiro}, {Gee}, {Guida}, {Reynolds}, \& {Walls}}]{Lares-Martiz2024}
{Lares-Martiz}, M., {Buzasi}, D., {Oswalt}, T., {et~al.} 2024, \bibinfo{title}{{How Reliable are Rotation Period Determinations from TESS Data?},} Research Notes of the American Astronomical Society, 8, 132, \dodoi{10.3847/2515-5172/ad4a7c}

\bibitem[{Y. {Lu} {et~al.}(2020){Lu}, {Angus}, {Ag{\"u}eros}, {Blancato}, {Ness}, {Rowland}, {Curtis}, \& {Grunblatt}}]{Lu2020}
{Lu}, Y., {Angus}, R., {Ag{\"u}eros}, M.~A., {et~al.} 2020, \bibinfo{title}{{Astraea: Predicting Long Rotation Periods with 27 Day Light Curves},} \aj, 160, 168, \dodoi{10.3847/1538-3881/abada4}

\bibitem[{T.~S. {Metcalfe} {et~al.}(2023){Metcalfe}, {Buzasi}, {Huber}, {Pinsonneault}, {van Saders}, {Ayres}, {Basu}, {Drake}, {Egeland}, {Kochukhov}, {Petit}, {Saar}, {See}, {Stassun}, {Li}, {Bedding}, {Breton}, {Finley}, {Garc{\'\i}a}, {Kjeldsen}, {Nielsen}, {Ong}, {R{\o}rsted}, {Stokholm}, {Winther}, {Clark}, {Godoy-Rivera}, {Ilyin}, {Strassmeier}, {Jeffers}, {Marsden}, {Vidotto}, {Baliunas}, \& {Soon}}]{Metcalfe2023}
{Metcalfe}, T.~S., {Buzasi}, D., {Huber}, D., {et~al.} 2023, \bibinfo{title}{{Asteroseismology and Spectropolarimetry of the Exoplanet Host Star {\ensuremath{\lambda}} Serpentis},} \aj, 166, 167, \dodoi{10.3847/1538-3881/acf1f7}

\bibitem[{M.~B. {Nielsen} {et~al.}(2020){Nielsen}, {Ball}, {Standing}, {Triaud}, {Buzasi}, {Carboneau}, {Stassun}, {Kane}, {Chaplin}, {Bellinger}, {Mosser}, {Roxburgh}, {{\c{C}}elik Orhan}, {Y{\i}ld{\i}z}, {{\"O}rtel}, {Vrard}, {Mazumdar}, {Ranadive}, {Deal}, {Davies}, {Campante}, {Garc{\'\i}a}, {Mathur}, {Gonz{\'a}lez-Cuesta}, \& {Serenelli}}]{Nielsen2020}
{Nielsen}, M.~B., {Ball}, W.~H., {Standing}, M.~R., {et~al.} 2020, \bibinfo{title}{{TESS asteroseismology of the known planet host star {\ensuremath{\lambda}}$^{2}$ Fornacis},} \aap, 641, A25, \dodoi{10.1051/0004-6361/202037461}

\bibitem[{B. {Nordstr{\"o}m} {et~al.}(2004){Nordstr{\"o}m}, {Mayor}, {Andersen}, {Holmberg}, {Pont}, {J{\o}rgensen}, {Olsen}, {Udry}, \& {Mowlavi}}]{Nordstrom2004}
{Nordstr{\"o}m}, B., {Mayor}, M., {Andersen}, J., {et~al.} 2004, \bibinfo{title}{{The Geneva-Copenhagen survey of the Solar neighbourhood. Ages, metallicities, and kinematic properties of {\ensuremath{\sim}}14 000 F and G dwarfs},} \aap, 418, 989, \dodoi{10.1051/0004-6361:20035959}

\bibitem[{T.~D. Oswalt {et~al.}(2022)Oswalt, Buzasi, Otani, Vaidya, Shanahan, \& Grigorov}]{Oswalt2022}
Oswalt, T.~D., Buzasi, D., Otani, T., {et~al.} 2022, \bibinfo{title}{Assessing Rotation Rates Among Components of Wide Binaries in the TESS and Kepler/K2 Surveys,} Zenodo, \dodoi{10.5281/zenodo.7406311}

\bibitem[{L.~M. {Rebull} {et~al.}(2016){Rebull}, {Stauffer}, {Bouvier}, {Cody}, {Hillenbrand}, {Soderblom}, {Valenti}, {Barrado}, {Bouy}, {Ciardi}, {Pinsonneault}, {Stassun}, {Micela}, {Aigrain}, {Vrba}, {Somers}, {Gillen}, \& {Collier Cameron}}]{Rebull2016}
{Rebull}, L.~M., {Stauffer}, J.~R., {Bouvier}, J., {et~al.} 2016, \bibinfo{title}{{Rotation in the Pleiades with K2. II. Multiperiod Stars},} \aj, 152, 114, \dodoi{10.3847/0004-6256/152/5/114}

\bibitem[{A.~R.~G. {Santos} {et~al.}(2021){Santos}, {Breton}, {Mathur}, \& {Garc{\'\i}a}}]{Santos2021}
{Santos}, A.~R.~G., {Breton}, S.~N., {Mathur}, S., \& {Garc{\'\i}a}, R.~A. 2021, \bibinfo{title}{{Surface Rotation and Photometric Activity for Kepler Targets. II. G and F Main-sequence Stars and Cool Subgiant Stars},} \apjs, 255, 17, \dodoi{10.3847/1538-4365/ac033f}

\bibitem[{A.~R.~G. {Santos} {et~al.}(2017){Santos}, {Cunha}, {Avelino}, {Garc{\'\i}a}, \& {Mathur}}]{santos2017}
{Santos}, A.~R.~G., {Cunha}, M.~S., {Avelino}, P.~P., {Garc{\'\i}a}, R.~A., \& {Mathur}, S. 2017, \bibinfo{title}{{Starspot signature on the light curve. Learning about the latitudinal distribution of spots},} \aap, 599, A1, \dodoi{10.1051/0004-6361/201629923}

\bibitem[{A.~R.~G. {Santos} {et~al.}(2019){Santos}, {Garc{\'\i}a}, {Mathur}, {Bugnet}, {van Saders}, {Metcalfe}, {Simonian}, \& {Pinsonneault}}]{Santos2019}
{Santos}, A.~R.~G., {Garc{\'\i}a}, R.~A., {Mathur}, S., {et~al.} 2019, \bibinfo{title}{{Surface Rotation and Photometric Activity for Kepler Targets. I. M and K Main-sequence Stars},} \apjs, 244, 21, \dodoi{10.3847/1538-4365/ab3b56}

\bibitem[{A. {Skumanich}(1972){Skumanich}}]{Skumanich1972}
{Skumanich}, A. 1972, \bibinfo{title}{{Time Scales for Ca II Emission Decay, Rotational Braking, and Lithium Depletion},} \apj, 171, 565, \dodoi{10.1086/151310}

\bibitem[{K.~G. {Stassun} {et~al.}(2018){Stassun}, {Oelkers}, {Pepper}, {Paegert}, {De Lee}, {Torres}, {Latham}, {Charpinet}, {Dressing}, {Huber}, {Kane}, {L{\'e}pine}, {Mann}, {Muirhead}, {Rojas-Ayala}, {Silvotti}, {Fleming}, {Levine}, \& {Plavchan}}]{Stassun2018}
{Stassun}, K.~G., {Oelkers}, R.~J., {Pepper}, J., {et~al.} 2018, \bibinfo{title}{{The TESS Input Catalog and Candidate Target List},} \aj, 156, 102, \dodoi{10.3847/1538-3881/aad050}

\bibitem[{P.~W. Sullivan {et~al.}(2015)Sullivan, Winn, Berta-Thompson, Charbonneau, Deming, Dressing, Latham, Levine, McCullough, Morton, Ricker, Vanderspek, \& Woods}]{Sullivan2015}
Sullivan, P.~W., Winn, J.~N., Berta-Thompson, Z.~K., {et~al.} 2015, \bibinfo{title}{THE TRANSITING EXOPLANET SURVEY SATELLITE: SIMULATIONS OF PLANET DETECTIONS AND ASTROPHYSICAL FALSE POSITIVES,} The Astrophysical Journal, 809, 77, \dodoi{10.1088/0004-637X/809/1/77}

\bibitem[{R. Vanderspek {et~al.}(2018)Vanderspek, Doty, Fausnaugh, \& et~al.}]{Vanderspek2018}
Vanderspek, R., Doty, J., Fausnaugh, M., \& et~al. 2018, \bibinfo{title}{TESS Instrument Handbook,}, v0.1

\end{thebibliography}
\bibliographystyle{aasjournal}



\end{document}